\documentclass[onecolumn,sort&compress,numbers]{els-mrw} 

\usepackage{amsmath,amssymb,amsfonts,amsthm,makeidx,graphicx}
\usepackage{txfonts}
\usepackage{helvet}
\usepackage{color}
\usepackage[colorlinks=true,allcolors=Purple,pdfusetitle]{hyperref}

\begin{document}


\chapter{Neutrinos from diffuse supernova background}
\label{sec:DSNB}

\author[1]{Anna M. Suliga}%

\address[1]{\orgname{New York University}, \orgdiv{Center for Cosmology and Particle Physic}, \orgaddress{New York, NY 10003, USA}}

\articletag{Chapter Article tagline: Neutrinos from diffuse supernova background}

\maketitle


\begin{abstract}[Abstract]
Neutrinos are the second most abundant particles in the universe according to the Standard Model, yet they are the least likely to interact. This feature implies that detecting a neutrino can reveal valuable insights into its source. Among the known sources of neutrinos, core-collapse supernovae are one of the most efficient factories. On average, a single collapse occurs every second in the observable universe, emitting approximately $10^{58}$ neutrinos. The total flux of neutrinos reaching Earth from all core-collapse supernovae across the universe is the diffuse supernova neutrino background (DSNB). Detection of the DSNB is just around the corner. This guaranteed flux of astrophysical neutrinos encodes information about the whole supernova population, including an answer to a currently unsolved question about the rate at which black holes form from massive stars. 
This chapter discusses the ingredients entering the DSNB calculation as well as current experimental limits and hints.
\end{abstract}

\begin{keywords}
 	neutrinos\sep supernovae\sep multimessenger astronomy 
\end{keywords}

\section{Introduction}\label{intro}

The night sky has always been a subject of human interest. Since the first observations with the naked eye, humans have developed a myriad of telescopes and examined the whole range of the electromagnetic spectrum. These observations allowed us to decode precious pieces of information about multitude of astrophysical objects. However, only in the last century have we entered a new era -- multimessenger astrophysics. One type of messengers -- neutrinos -- are among the most intriguing couriers.

The perverse nature of neutrinos arises from the fact that, although they interact only weakly, they are the second most abundant Standard Model (SM) particles in our Universe. Their nature also allows them to directly probe extreme environments without losing information on their way to the Earth as photons or charged particles do. Consequently, neutrinos can provide insights beyond those that photons and charged particles carry. One of the most efficient, ubiquitous, and explosive neutrino factories are core-collapse supernovae (CCSN)~\citep{Colgate:1966ax, Bethe:1985sox, Bruenn:2012mj, Takiwaki:2013cqa, OConnor:2018sti, Janka:2016fox, Mezzacappa:2020oyq, Burrows:2020qrp}. A single collapse process in a span of about 10 seconds produces around $10^{58}$ neutrinos, where each is carrying energy of approximately 10-15 MeV. It is over 18 orders of magnitude higher luminosity than the one of the Sun in photons. The Nobel prize-winning detection of neutrinos from Supernova 1987A was a huge milestone in the fields of both particle physics and astrophysics. For the latter, this measurement confirmed that the explosions of massive stars are the result of a collapse of their cores happening at the end of their life, once they have gone through the multiple fuel-burning phases. As for the particle physics milestone, the observation of the neutrinos from the supernova 1987A set the best at the time limits on the absolute neutrino mass.

Today, the supernova neutrinos are again entering the spotlight. The Super-Kamiokande neutrino detector starts to see hints~\citep{SK-slides-2024} of the diffuse supernova neutrino backgroud (DSNB) -- a guaranteed neutrino flux that encodes information about the entire CCSN population in the observable universe~\citep{1984NYASA.422..319B, Krauss:1983zn, Wilson:1986ha, Ando:2004hc} (for more recent reviews see, e.g.~\citep{Beacom:2010kk, Lunardini:2010ab, Kresse:2020nto, Horiuchi:2020jnc, Ekanger:2023qzw, Suliga:2022ica}). Its detection will serve as an independent indirect measurement of the CCSN rate, the fraction of black-hole-forming supernova progenitors, the average neutrino flux per progenitor, and if we are lucky any potential signatures of physics beyond the SM.

The rest of this chapter is organized as follows: Section~\ref{sec:DSNB} briefly outlines the ingredients necessary for the DSNB calculation and discusses the uncertainties of the astrophysical parameters entering the estimation. Section~\ref{sec:Detectors} discusses the current experimental limits on the DSNB and the future sensitivities.

\section{Diffuse supernova neutrino background}
\label{sec:DSNB}

The minimal prediction of the DSNB relies on understanding the average neutrino emission per CCSN and the expected rate of such collapses in the universe. The former is addressed in Section~\ref{sed:flux}, while the latter is discussed in Section~\ref{sec:CC-rate}. Section~\ref{sec:DSNBplots} illustrates the range of the DSNB predictions.

\subsection{Average neutrino emission per core-collapse}
\label{sed:flux}

Core-collapse supernovae (CCSNe) are one of the most powerful neutrino sources~\citep{ALEXEYEV1988209, PhysRevLett.58.1490, Bionta:1987qt}. 
The extreme conditions in the centers of these massive stars enable efficient production of large amounts of neutrinos through electron capture on free protons and nuclei~\citep{Bethe:1979zd, Fuller:1981mv, Fuller:1981mu, 1985ApJS...58..771B, Martinez-Pinedo:2012eaj}, as well as deexcitation and thermal reactions of nuclei~\citep{1991ApJ...376..701F, Fischer:2013qpa, Martinez-Pinedo:2012eaj,Sullivan:2015kva}. 
The neutrinos created in such a way take away nearly the entire gravitational binding energy of the star over a period of only tens of seconds~\citep{Burrows:1986me}. The detailed description of the CCSN neutrino emission is a subject of another chapter in this Encyclopedia. In this chapter, we provide a brief and simple summary to explain how the concept of the average neutrino flux emitted per CCSN.

The core-collapse SN explosion process starts with the infall of the inner stellar core. During this phase, the core, which has reached sufficiently high densities, begins to collapse rapidly due to the increasing rates of photodisintegration and electron captures on heavy nuclei.
As a result, the emission of neutrinos of all flavors, but especially electron neutrinos, grows exponentially. 
However, once the density of matter increases to about $10^{12}\;\mathrm{g}\;\mathrm{cm}^{-3}$, the neutrino diffusion timescale surpasses the collapse timescale. Thus, neutrinos become trapped in the core, and further neutrino emission is controlled by flavor-dependent and independent scattering reactions with the matter~\cite{Raffelt:2001kv, Fischer:2011cy}. 
Once the density in the core reaches the nuclear saturation density of approximately $2.3 \times 10^{14}~\mathrm{g}\;\mathrm{cm}^{-3}$, the collapse is stopped by the repulsive short-range nuclear forces. The core bounces, and a shock wave is launched from the inner/outer core boundary leaving protoneutron star in the center.  Initially, it was assumed that the bounce shock was energetic enough to promptly travel across the star and result in a visible photon explosion. However, detailed numerical simulations have shown that the bounce shock loses energy as it propagates and stalls after traveling a few hundred kilometers. Refs.~\citep{Colgate:1966ax} and~\citep{Bethe:1985sox} postulated and explained how neutrinos oozing from the inner core could interact within the shock and reenergize it, leading to a supernova explosion. The modern state-of-the-art 3D simulations~\citep{Burrows:2000mk, Janka:2012wk, Ott:2012mr, Lentz:2015nxa, Janka:2016fox, Takiwaki:2016qgc, Vartanyan:2018iah, Burrows:2020qrp, Kuroda:2020bdq} demonstrate that the delayed neutrino heating mechanism aided by the hydrodynamical instabilities and convection-driven turbulence can lead to successful explosions~\citep{Blondin:2002sm, Tamborra:2014aua, Couch:2014kza}. After the explosion, the protoneutron star left in the center of the star cools down by neutrino emission and forms a neutron star.

The distribution of neutrino energies emitted by the core-collapse, on average, rather closely follows a thermal spectrum. The fit best describing the shape of that spectrum is called pinched thermal spectrum~\citep{Keil:2003sw, Keil:2002in, Tamborra:2012ac} and takes the form
\begin{equation}
\label{eq:spectrum}
	\varphi_{\nu_i}(E_{\nu_i}, t) = \xi_{\nu_i}(t) \left( \frac{E_{\nu_i}}{\langle E_{\nu_i}(t)\rangle}\right)^{\alpha_{\nu_i}(t)} \exp \left(-\frac{(\alpha_{\nu_i}(t)+1)E_{\nu_i}}{\langle E_{\nu_i}(t)\rangle}\right) \ ,
\end{equation}
where $E_{\nu_i}$ is the energy of the neutrino flavor $i=\{e, \mu, \tau\}$, $\langle E_{i}(t)\rangle$ is the average energy, $t$ is the time of the neutrino emission, and the normalization factor $\xi_{\nu_i}(t) = (\int dE_{\nu_i} \varphi_{\nu_i}(E_{\nu_i}, t))^{-1}$.
The pinching parameter $\alpha_{i}(t)$ represents how much the spectrum differs from a pure Fermi-Dirac distribution, which is recovered for $\alpha_{i}(t) \equiv \alpha = 2.3$. In that case, it is also common to connect the mean energy of neutrinos to their temperature by the expression $\langle E_{i}(t)\rangle = 3.15 \; T_{\nu_i} (t)$. 
The differential flux of neutrinos ($\nu_i = \{\nu_{e},\bar{\nu}_{e},\nu_{\mu},\bar{\nu}_{\mu}, \nu_{\tau},\bar{\nu}_{\tau} \}$) can be then expressed by the relation
\begin{equation}
\label{eq:flux}
	f_{\nu_i} (E_{\nu_i}, t) = \frac{L_{\nu_i}(t)}{\langle E_{\nu_i}(t)\rangle}\frac{\varphi_{\nu_i}(E_{\nu_i}, t)}{4\pi D^2}  = \frac{F_{\nu_i} (E_{\nu_i}, t)}{4\pi D^2} \ ,
\end{equation}
where $L_{i} (t)$ is the neutrino luminosity and $D$ is the distance to the supernova. For the nearby CCSN detection, it is critical to measure the temporal evolution of the neutrino signal. It can reveal information about how the core temperature is changing~\citep{Horiuchi:2018ofe}, the localization of the supernova progenitor~\citep{Beacom:1998fj, Muhlbeier:2013gwa, Brdar:2018zds, Hansen:2019giq, Linzer:2019, Coleiro:2020vyj, Sarfati:2022}, allow us to set a limit on the absolute neutrino mass~\citep{Zatsepin:1968kt, Totani:1998nf, Beacom:2000ng, Loredo:2001rx, Nardi:2003pr, Nardi:2004zg, Pagliaroli:2010ik, Lu:2014zma, Rossi-Torres:2015rla, Hansen:2019giq, Gullin:2021hfv, Pompa:2022cxc, Pitik:2022jjh}, reveal if a phase transition between hadrons and quarks took place~\citep{Migdal:1979je, Takahara:1988yd, Gentile:1993ma, Pons:2001ar, Drago:2008tb, Nakazato:2008su, Sagert:2008ka, Fischer:2010wp, Fischer:2017lag, Zha:2021fbi, Fischer:2021tvv, Kuroda:2021eiv, Bauswein:2022vtq, Lin:2022lck, Jakobus:2022ucs, Pitik:2022jjh}, or if there are imprints of physics beyond the SM in the signal, see, e.g., Refs~\citep{Suliga:2020jfa, Lazar:2024ovc, Suliga:2024nng, Beatty:2025bgq}. On the other hand, DSNB detection requires a long experimental exposure to see the actual flux. Therefore, only the total time-integrated or average flux emitted from a single CCSN matters.

The core collapse of a massive star can also, instead of leading to an explosion and leaving a neutron star as a remnant, lead to direct black hole formation~\citep{Fryer:1999mi, Sekiguchi:2004ba, Sekiguchi:2004tma, Zhang:2007nw, Fischer:2008rh, OConnor:2010moj}. Such an event can plausibly occur without any observable electromagnetic signal~\citep{Lovegrove:2013ssa}. In these cases, neutrinos may be the only detectable messengers from such an event. 
The stars directly collapsing to black holes are expected to produce more energetic neutrino spectra than standard neutron-star-forming CCSNe ~\citep{Lunardini:2009ya}.
This is because proto-neutron stars in black hole-forming progenitors can reach larger masses than those in standard CCSNe, creating hotter and denser environments. In the case of neutron-star-forming CCSN simulations, the mean energies of neutrinos tend to be between 10-18~MeV~\citep{Mirizzi:2015eza, OConnor:2018sti, Vartanyan:2023zlb}, depending on the flavor. In the case of black-hole-forming models, the mean energies can grow to more than approximately 20~MeV~\citep{Mirizzi:2015eza, OConnor:2018sti,Vartanyan:2023zlb}.

\subsection{Core-collapse frequency in the Universe}
\label{sec:CC-rate}

Core-collapse occurs for relatively massive stars ($8~M_\odot \lesssim M \lesssim 125~M_\odot$)~\citep{Heger:2002by} which are characterized by short lifetimes (approximately $10^6$~yr) on a cosmic scale. Due to this, the frequency of core-collapse events can be calculated directly using the star formation rate (SFR, $\rho_*$) and initial mass function (IMF, $\xi$). These two parameters tell respectively about the speed of the formation of stars per co-moving volume and the distribution of stars' masses at their births. Using them, the core-collapse supernova rate (CCSNR, $R_\mathrm{CC}$) can be calculated as
\begin{equation}
\label{eq:R_CC}
R_\mathrm{CC} (z, M) = \frac{\dot\rho_{*}(z) \xi}{\int_{0.1~M_\odot}^{125} dM M \xi} \ .
\end{equation}
Employing this method means that any uncertainties impacting the measurements of the SFR and IMF will propagate to the CCSNR, and consequantly DSNB; see, e.g., \citep{Horiuchi:2008jz, Lunardini:2012ne, Mathews:2014qba, 2011ApJ...738..154H, Moller:2018kpn, Singh:2020tmt, Suliga:2021hek, Ziegler:2022ivq}.

Another way of determining the core-collapse rate is by identifying the frequency of the supernova events using electromagnetic observations~\citep{Mattila:2012zr, Dahlen:2012cm, Strolger:2015kra, Petrushevska:2016kie}. This method also has its challenges. One of them is the fact that these are low-frequency events at small redshifts~\citep{Beacom:2010kk}. Moreover, the observations can underestimate the rate if not all core-collapse lead to an observable explosion, for example, when the star instead evolves directly into a black hole or a significant dust obscuration prevents clean observation~\citep{Mattila:2012zr}. 
In addition, Ref.~\citep{2011ApJ...738..154H} identified a tension by a factor of two between the supernova rate calculated using the IMF and SFR and the one obtained directly from the observations of core-collapse explosions. Fortunately, in the coming years, significantly more accurate core-collapse rate estimations can be made using electromagnetic observation data from the already operational Vera Rubin Observatory~\citep{LSST:2008ijt, Lien:2009db, Ekanger:2023qzw}. This telescope is expected to observe an unprecedented number of supernova events in a short period, covering nearly half of the sky. With a planned revisit time of about three days per quarter of the sky and two visits per night, such a survey will lead to an order-of-magnitude increase in the number of supernovae discovered.

The percentage of progenitors evolving into black holes is a major fundamental unknown in astrophysics. Determining this parameter is crucial for understanding the detailed physics of CCSN simulations and the origins of the observed binary black hole mergers detected by the LIGO-Virgo-KAGRA collaborations~\citep{LIGOScientific:2025slb}. Direct observations of such events are very challenging. So far, only two candidates for disappearing red giants have been observed with the Large Binocular Telescope~\citep{Kochanek:2008mp, Lien:2010yb, Gerke:2014ooa, Adams:2016ffj, Adams:2016hit, Davies:2020iom, Neustadt:2021jjt}. One of them was followed up with the James Webb Space Telescope~\citep{Beasor:2023ltg}, though it still remains unclear whether it is a true direct collapse to a black hole event. There is also indirect evidence of a star directly collapsing into a black hole in a binary system observed by Gaia telescope. The system consists of a massive electromagnetically invisible component, presumably a black hole, and a red giant star in a nearly circular orbit~\citep{Regaly:2025xnc}. Such an orbital configuration is generally thought to be unlikely to survive a full-blown supernova explosion, unless a finely-tuned explosion occurs when the secondary is at the farthest point from the center of attraction of the binary system. Population studies using one-dimensional supernova simulations of the expected fraction of black-hole-forming CCSN have been conducted, yielding results that vary significantly between a few to tens of percent~\citep{OConnor:2011pxx, Sukhbold:2015wba, Sukhbold:2013yca, Ertl:2015rga, Couch:2020}.

\subsection{Diffuse supernova neutrino background estimates}
\label{sec:DSNBplots}

The diffuse supernova background summed over all neutrino flavors can be calculated with the expression
\begin{equation}
\label{eq:DSNB}
    \Phi (E_\nu) = \frac{c}{H_0}\int_0^{z_\mathrm{max}} dz \frac{1}{\sqrt{\Omega_\mathrm{M}(1+z)^3 + \Omega_\Lambda}} \int_{8M_\odot}^{125M_\odot} dM \; F_{\nu}(E_\nu^\prime, M) \; R_\mathrm{CC} (z, M) \ ,
\end{equation}
where $c$ is the speed of light, $H_0$ is the Hubble constant, $z$ is the cosmological redshift, $E_\nu^\prime=E_\nu(1+z)$, $\Omega_M$ and $\Omega_\Lambda$ are respectively the fractions of the energy density of the Universe in matter and dark energy, and $F(E_\nu^\prime, M)$ is the time-integrated flux from a single CCSN in all flavors.

\begin{figure}[t]
\centering
\includegraphics[width=8cm]{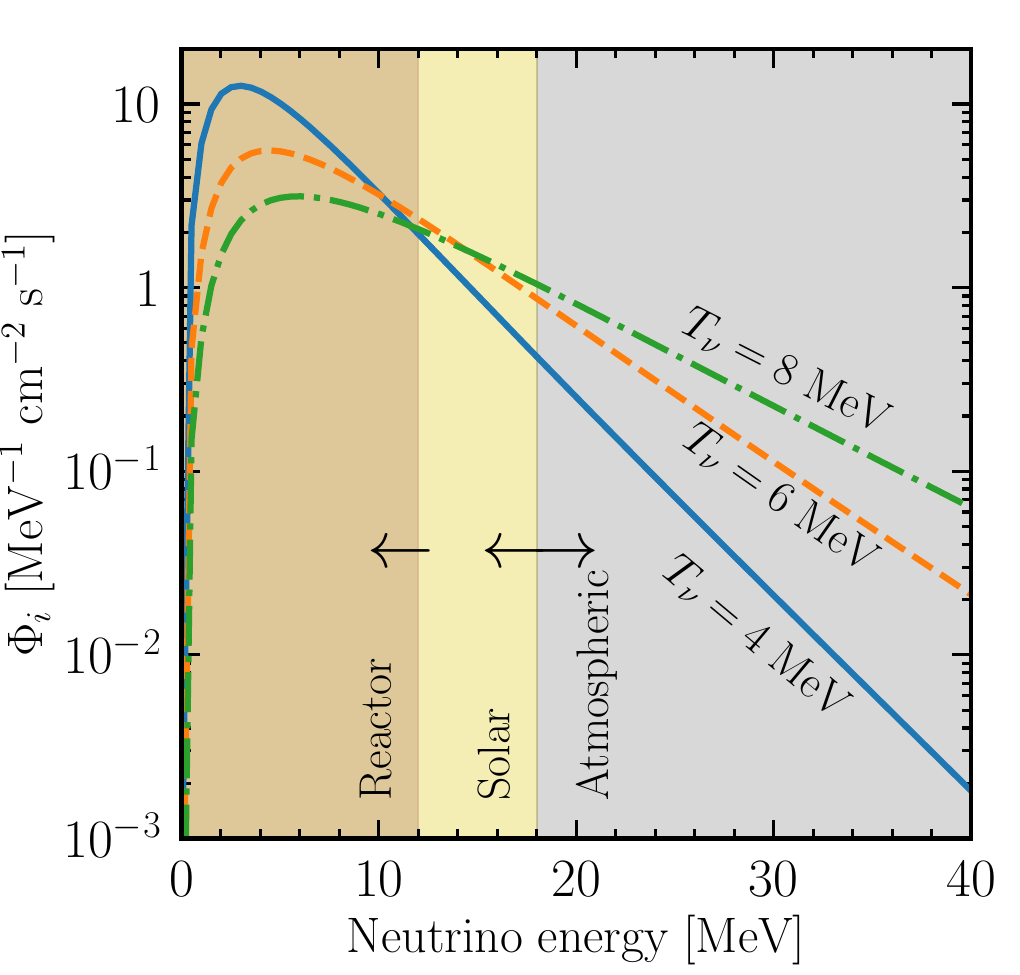}
\caption{The DSNB for sum of all flavors assuming the Fermi-Dirac spectrum described by the temperature $T_\nu$ and the total energy emitted in all flavors $3\times 10^{53}$~erg. Different $T_{\nu}$ reflect how one source of uncertainty - the spectrum emitted from CCSN - can modify the DSNB. The shaded bands labeled by the Reactor, Solar, and Atmospheric mark the regions where these three nonreducible backgrounds affect the DSNB measurement.}
\label{fig:DSNB}
\end{figure}

Figure~\ref{fig:DSNB} shows the DSNB estimated using~\citep{1955ApJ...121..161S}~IMF, nominal SFR taken from~\citep{2011ApJ...738..154H}, and assuming that all the CCSNe emit the Fermi-Dirac neutrino spectrum characterized by the temperature $T_\nu$ and total energy emitted by all flavors is $3\times 10^{53}$~erg. The figure highlights how one of the uncertainties - the shape of the neutrino spectrum emitted from CCSN - modifies the DSNB. It also points out that higher fraction-of-black-hole forming stars in the entire CCSN population increases the high energy tail of the DSNB~\citep{Lunardini:2009ya}, as these stars are expected to produce hotter spectra.
The shaded bands labeled Reactor, Solar, and Atmospheric indicate in which energy regions these irreducible neutrino backgrounds affect the DSNB measurement (see~Section~\ref{sec:Detectors}). 

Other factors, not discussed above, that can influence the shape and normalization of the DSNB include neutrino flavor conversions within the supernova and the effects of binary star evolution. We will briefly discuss the impact of both of these effects on the DSNB in the next two paragraphs. 

While propagating through the supernova medium, neutrinos undergo flavor conversions due to their self-interactions and coherent forward scattering of the medium particles. The neutrino conversions may affect the DSNB flavor content. The solution to the neutrino flavor evolution inside the core-collapse supernova is yet to be found. The main difficulty, compared to neutrino flavor evolution in the Sun, is the possibility of the neutrino self-interactions which may lead to highly non-linear flavor evolution~\citep{Duan:2010bg, Balantekin:2006tg, Chakraborty:2016yeg, Tamborra:2020cul}. Thus, detecting the DSNB in all flavors is vital to disentangle the astrophysical uncertainties from the effect of neutrino flavor conversions. 

Observational evidence points out that most stars reside in binary systems~\citep{Sana:2012px, Zapartas:2020cdi}, which opens up the possibility of binary interactions and introduces additional uncertainty in the DSNB. The binary interactions of massive stars can lead to redistribution of masses in the supernova population due to the mass transfer or changes in the number of stars undergoing core collapses by mergers. If two progenitors that were not expected to undergo supernovae form a star massive enough to undergo core collapse, the number of CCSN increases. On the other hand, if both stars before the merger were expected to undergo CCSN, the mergers decreased the number of CCSN. Convolving these arguments with the IMF leads to the conclusion that the number of CCSN after including binary interactions increases. The effect of binary interactions on the DSNB can lead to a 0-75\% increase in the flux~\citep{Horiuchi:2020jnc}. The exact details depend on the modeling of the common envelope, in particular, how easy it is to unbind it and whether rotation effects are taken into account. The latter can lead to the development of more massive cores, which result in a larger neutrino flux~\citep{Horiuchi:2020jnc}.

It is hopefully now clear that the DSNB encodes valuable information about a wide range of astrophysical parameters. As a result, future DSNB measurements will provide an independent and indirect probe of the supernova rate, the fraction of black-hole-forming progenitors, and other key astrophysical factors~\citep{Lunardini:2009ya, Keehn:2010pn, 2011ApJ...738..154H, Nakazato:2013maa, Nakazato:2015rya, Priya:2017bmm, Horiuchi:2017qja, Moller:2018kpn, Kresse:2020nto, Singh:2020tmt, Horiuchi:2020jnc, Ziegler:2022ivq, Libanov:2022yta, Ekanger:2022neg, Ashida:2023heb, Nakazato:2024gem}.

\section{Detection of the diffuse supernova neutrino background}
\label{sec:Detectors}

The fact that neutrinos interact only by the weak force is a double-edged sword. On the one hand, neutrinos can travel vast distances unabsorbed; however, their detection in terrestrial experiments is a major challenge. 
A key way of mitigating the hardship of measuring astrophysical neutrinos is constructing large detectors. Their sizeable volumes guarantee many interaction targets. The second crucial step is shielding these detectors from numerous particles entering their volume and interacting with strong and electromagnetic forces. 
In this section, we describe DSNB detection prospects in a variety of neutrino experiments.

\subsection{Detection of the electron antineutrino component}

The diffuse supernova neutrino background has not been observed so far. But Super-Kamiokande (SK), one of the existing large-scale water Cherenkov detectors, is the leader in setting the upper limits on the $\bar\nu_e$ component of the DSNB. The latest results without the gadolinium loading, from the DSNB search at SK, indicated that $\Phi_{\bar\nu_e} < 2.7$ cm$^{-2}$~s$^{-1}$, for $\bar\nu_e$ energies larger than 17.3 MeV \citep{Bays:2011si, Super-Kamiokande:2013ufi, Super-Kamiokande:2021jaq}.  
The results from the search with gadolinium loaded water already achieved the same level of sensitivity~\citep{Super-Kamiokande:2023xup} with approximetly five times shorter exposure.

\begin{figure}[t]
\centering
\includegraphics[scale=0.35]{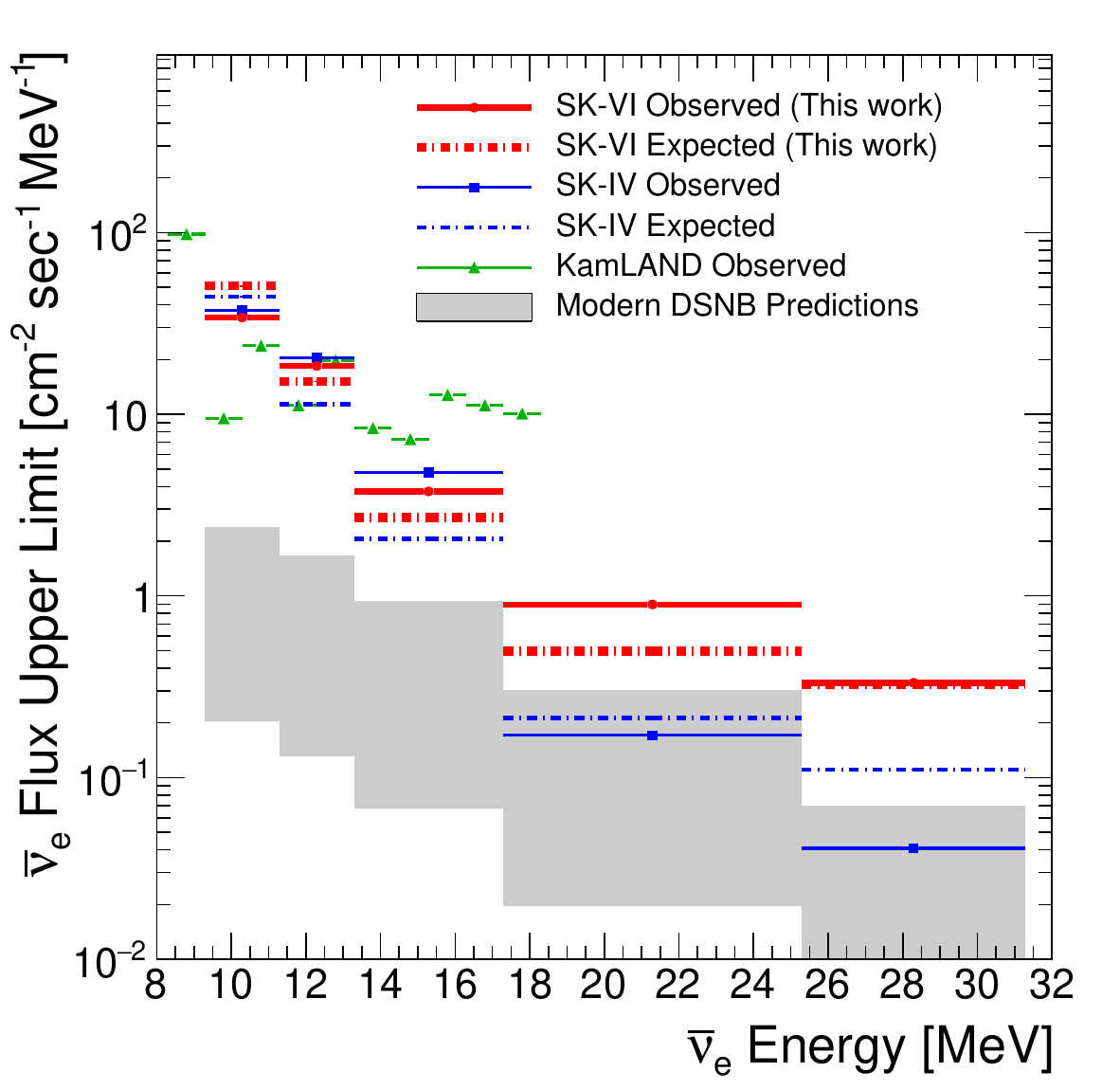}
\caption{Upper limits on the $\bar\nu_e$ component of the DSNB from SK-IV~\cite{Super-Kamiokande:2021jaq}, SK-VI (first gadolinium loaded results)~\cite{Super-Kamiokande:2023xup}, and KamLand~\cite{KamLAND:2021gvi} (colored markers) together with theoretical predictions (gray lines). Figure extracted from Ref.~\cite{Super-Kamiokande:2023xup}.}
\label{fig:Limits-barnue}
\end{figure}

\subsubsection{Water Cherenkov detectors}
\label{sec:water_cher_det}

Water Cherenkov detectors operate utilizing the fact that charged particles that move at speeds higher than the speed of light of the detector medium emanate Cherenkov radiation. The detector then registers this light by photomultiplier tubes and detects the moving particle.  But since neutrinos do not have an electric charge, to register their presence in the detector, they need to interact with the detector targets either by the momentum transfer or creation of a charged particle. In water, for neutrinos with energies close to $\mathcal{O}(10)$~MeV, $\bar \nu_e$ have the largest probability of interaction with free protons through the Inverse Beta Decay (IBD) 
\begin{equation}
\label{eq:IBD}
\bar \nu_e + p \rightarrow n + e^+ \ .
\end{equation}
Ultimately, because the number of the hydrogen (proton) targets surpasses twice the number of the oxygen targets, the threshold for IBD reaction is low ($E_\mathrm{th} = 1.806$ MeV), and both products of the reaction can be identified IBD is the primary DSNB observation channel in water Cherenkov detectors.

To reduce the backgrounds preventing the DNSB detection, SK has been adding gadolinium sulfate (GdCl${}_3$)~\citep{Beacom:2003nk} to the water in the tank since 2020. To understand how the addition of $\mathrm{GdCl}_3$ will help the background suppression, let us take a closer look at how SK identifies the IBD events.
In an IBD reaction, two products are made: a neutron and a positron. The positron emits Cherenkov radiation by which it can be identified. Unfortunately, in the DSNB energy window, several backgrounds emit Cherenkov radiation, e.g., decay electrons from invisible (with energies below the Chereknov threshold) atmospheric muons. To distinguish the signal from background events, coincident detection of neutron and positron is required.

A way of doing so is looking for Cherenkov radiation followed up shortly by $\gamma$ cascades, which come from excited by neutrons nuclei. This neutron-tagging procedure demands a target element that is both efficient neutron capturer and deexcites by emission of well-measured and observable $\gamma$ cascade. So far this has been done using free protons. But it turns out that gadolinium has a cross section for thermal neutron capture, 
\begin{equation}
\label{eq:Gd-deexcites}
\mathrm{n + Gd} \rightarrow \mathrm{Gd}^* \rightarrow  \mathrm{Gd} + \gamma \ ,
\end{equation}
$\approx 4.9 \cdot 10^4$ barn \citep{Gd-barn} (it is $10^5$ times higher than thermal capture on free protons) and deexcites fast by a $\gamma$ cascade with the total energy of 8~MeV which is much easier to observe in SK than 2.2~MeV deexcitation from protons. Addition of 0.1\% $\mathrm{GdCl}_3$ to the water in SK should reduce the invisible muon background by almost a factor of 5 and leave only single spallation product ($^9$Li) as background for the DSNB search~\citep{Beacom:2003nk, Super-Kamiokande:2021jaq}. SK enriched with Gd might provide $3\sigma$ detection for the DSNB detection already after 5-10 years~\citep{Beacom:2003nk}.
As of now the gadolinium enriched SK has already started seeing over 2$\sigma$ hints of that signal~\cite{SK-slides-2024}.

\subsubsection{Liquid scintillator detectors}
\label{sec:liquid-scintiallator-detector}

Another class of detectors which may measure the $\bar\nu_e$ component of the DSNB are the liquid scintillator detectors. This type of detectors can identify the IBD interaction by detecting both the prompt scintillation light coming from positron anihilation and its kinetic energy with coincidence by the delayed $\gamma$ rays from the neutron capture on proton.

Jiangmen Underground Neutrino Observatory (JUNO)~\citep{JUNO:2022hxd}, under construction, and THEIA~\citep{Theia:2019non}, proposed, are the large-scale detectors which aim to measure the $\bar\nu_e$ component of DSNB within the next decades. Using the pulse shape discrimination techniques~\citep{Mollenberg:2014pwa, Dunger:2019dfo}, these detectors are expected to suppress large neutral current atmospheric neutrino backgrounds and leave only the reactor and low energy atmospheric $\bar\nu_e$ fluxes as the backgrounds sources in the DSNB search window. JUNO is expecting to provide $3\sigma$ evidence for the DSNB detection after approximately 10 years of running~\citep{JUNO:2022lpc} and THEIA after 1 year thanks to its large volume~\citep{Theia:2019non}. 
THEIA would use a novel water-based liquid scintillator to get both scintillation and Cherenkov light, aiming to be the best of both worlds.

\subsection{Detection of the electron neutrino component}

Water Cherenkov and scintillator detectors cannot detect the $\nu_e$ component of the DSNB due to lower cross sections and high backgrounds present in the relevant energy range. Currently, the best upper limit on the $\nu_e$ component of the DSNB comes from the Sudbury Neutrino Observatory (SNO), which employed heavy water. The upper limit placed by SNO is $\Phi_{\nu_e} < 19$ cm$^{-2}$~s$^{-1}$, for $\nu_e$ energies between 22.9~MeV and 36.9~MeV~\citep{SNO:2020gqd}, which is approximately a factor of 20 above the majority of the theoretical predictions. There are multiple challenges that made the measurement nonfeasible, including the large solar neutrino rate and decaying muon background.

\begin{figure}[t]
\centering
\includegraphics[scale=0.35]{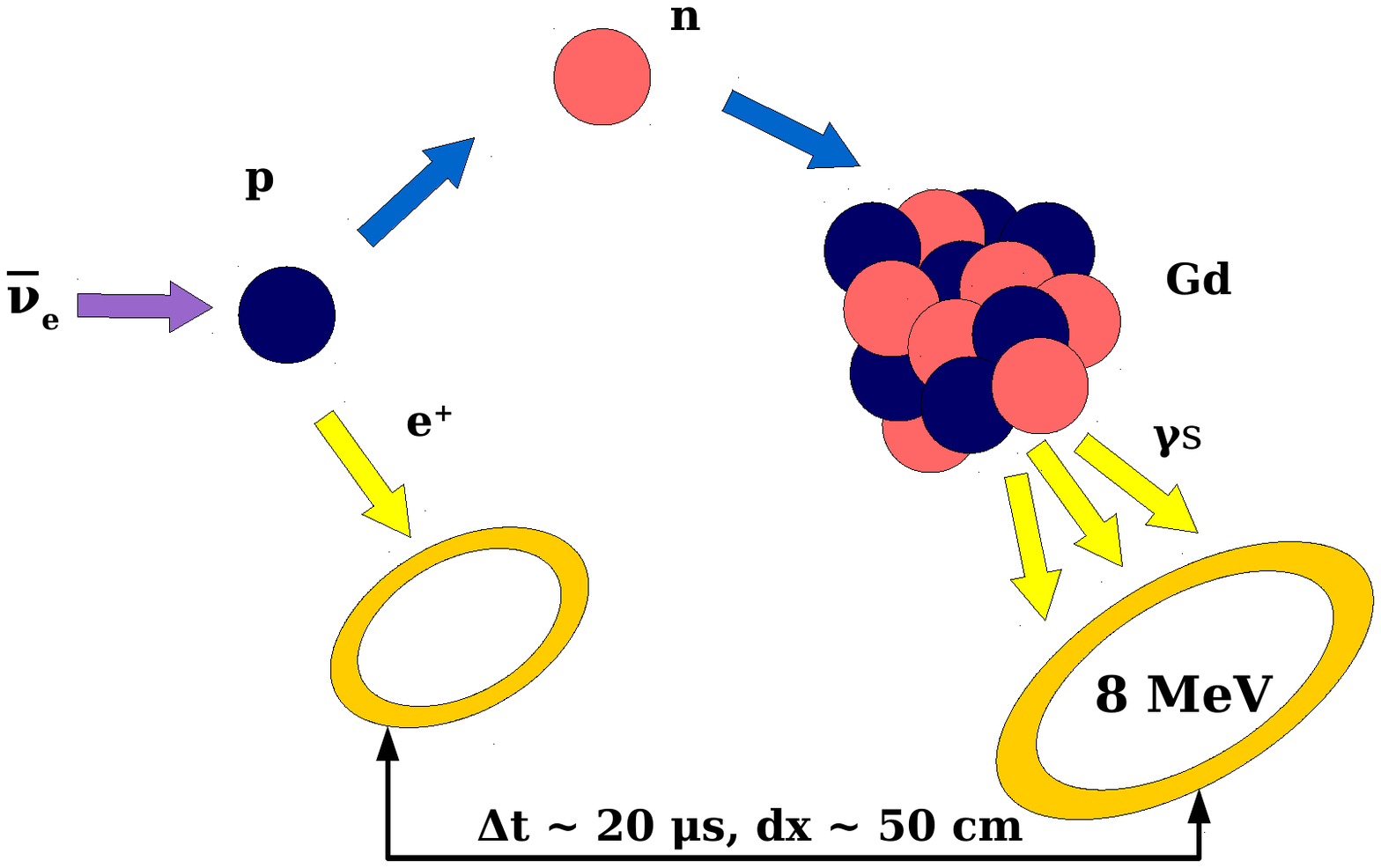}
\caption{Schematic of tagging the inverse beta decay in water with dissolved gadolinium sulfate (Figure based on the one from Ref.~\citep{Beacom:2003nk}).}
\label{IBD-GD}
\end{figure}

\subsubsection{Time projection chamber - Deep Underground Neutrino Experiment}
\label{sec:DUNE}

The Deep Underground Neutrino Experiment (DUNE) is a planned neutrino experiment that will include a large-scale Liquid Argon Time Projection Chamber (40 ktons fiducial volume)~\citep{DUNE:2021tad}. In this type of detector the main detection channel of the DSNB neutrinos is the charge-current interaction of $\nu_e$ with argon targets ($\nu_e + ^{40}\mathrm{Ar} \rightarrow e^- + ^{40}\mathrm{K}^*$). The large sensitivity to electron neutrinos is a feature that distinguishes DUNE from the water Cherenkov and liquid scintillator detectors; it introduces a new channel of observation for the DSNB.
 
The $\nu_e$ is identified by its interaction products, in a similar manner as $\bar\nu_e$ in water Cherenkov and scintillator detectors. The detector registers the ionization track and the scintillation light coming from the electron together with the $\gamma$ cascade from the deexcitation of potassium nuclei. If the spallation backgrounds can be efficiently reduced~\citep{Zhu:2018rwc} and DUNE operates on tens of years timescale, there is a possibility of detecting DSNB. In optimistic scenarios, $3\sigma$ could be achieved after 10 years of running~\citep{Moller:2018kpn}.

\subsection{Detection of the non-electron neutrino component}

While the prospects for detecting the $\nu_e$ and $\bar\nu_e$ components of the DSNB seem optimistic, to fully capture all the physics and astrophysics involved with the DSNB the measurement of the non-electron flavors ($\nu_\mu, \bar\nu_\mu, \nu_\tau, \bar\nu_\tau$ - that are commonly described as $\nu_x$ in supernova due to their similarity) is also necessary.
From the perspective of bare fluxes, the situation seems similar to $\nu_e$ and $\bar\nu_e$; the fluxes are comparable, and the detection window is nearly the same. Detecting MeV-scale non-electron neutrinos, however, is challenging because they cannot produce a charged lepton in their interactions as the lowest thresholds for the charged-current reactions are at least as high as the charged lepton masses. Due to this, the detection of the $\nu_x$ component of DSNB requires neutral-current channels.

A potential detection channel could be the elastic scattering of neutrinos on electrons, but it is not feasible since the signal cannot be distinguished from the ample backgrounds without the possibility to tag both products of the reactions in the detectors described in previous sections. The existing upper limit on the $\nu_x$ component of the DSNB is a thousand times worse than most of the theoretical models~\citep{Lunardini:2008xd}.

\subsubsection{Direct dark matter detectors - coherent elastic neutrino-nucleus scattering detectors}
\label{sec:CEvNS}
 
\begin{figure}[t]
\centering
\includegraphics[scale=0.5]{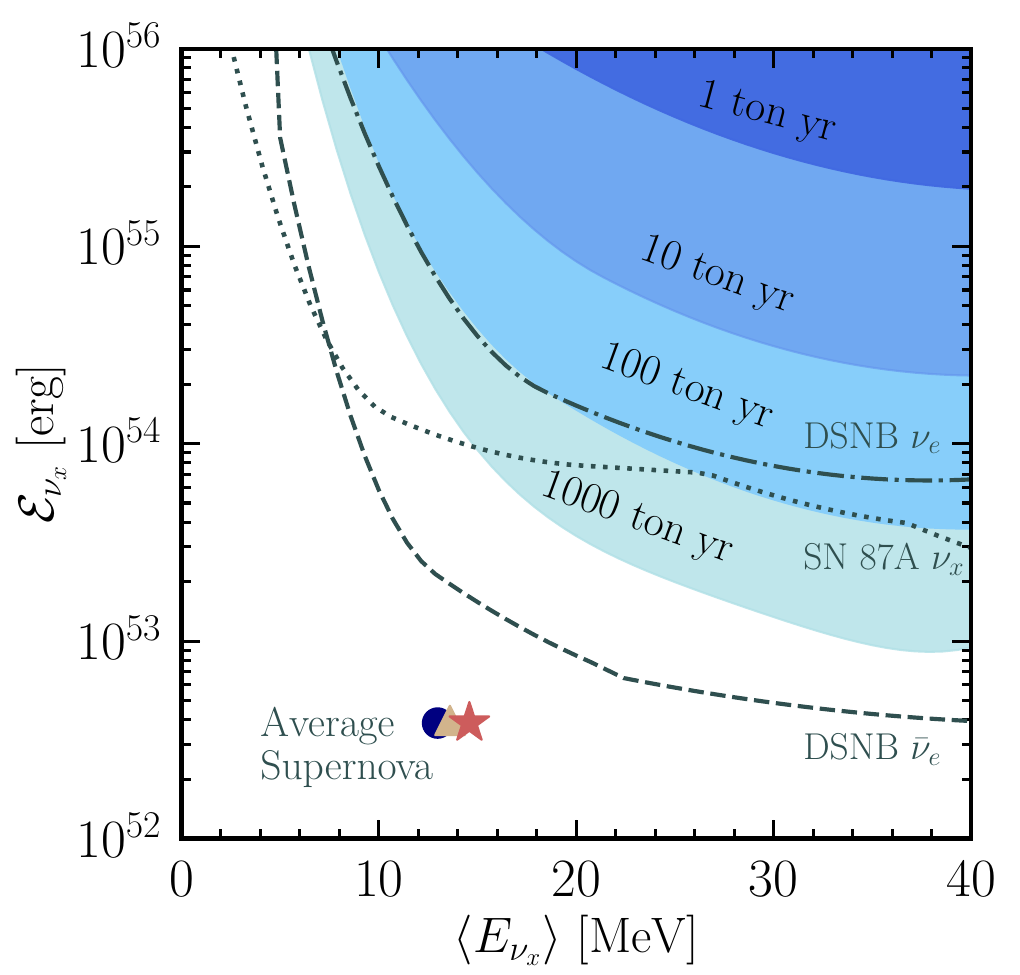}
\caption{The calculated sensitivity to the non-electron neutrino component of the DSNB in xenon-based CE$\nu$NS detectors. The y-axis ($\mathcal{E}_{\nu_x}$) is the total energy emitted by one non-electron neutrino flavor, whereas the x-axis ($\langle E_{\nu_x}\rangle$) shows the average neutrino energy. In addition, the current SK limit on $\bar\nu_e$~\citep{Super-Kamiokande:2021jaq} and the SNO limit on $\nu_e$~\citep{SNO:2020gqd}, the
SN 1987A limit on $\nu_x$~\citep{Suliga:2021hek}, and the average emission per collapse in nominal theoretical DSNB models~\citep{Moller:2018kpn} are shown.}
\label{fig:DSNB-limits-nux}
\end{figure}

Another possibility for observing non-electron neutrino component of the DSNB is the recently opened channel of neutrino detection -- the coherent elastic neutrino-nucleus scattering (CE$\nu$NS)~\citep{Akimov:2017ade}. This process has nearly the same cross section for all neutrino flavors~\citep{Freedman:1973yd}, and the coherence condition is satisfied within the detection window of DSNB for most nuclei employed by the CE$\nu$NS experiments. The detectors solely focused on measuring the neutrinos from accelerators~\citep{Akimov:2017ade}, and nuclear reactors~\citep{Akimov:2012aya, Agnolet:2016zir, Strauss:2017cuu, Hakenmuller:2019ecb, Aguilar-Arevalo:2019jlr} are too small to register even a single DSNB event over tens of years. However, large-scale direct dark matter detectors such as XENON \& LZ, DARWIN, PANDA~\citep{Aalbers:2016jon, XENON:2017lvq, LUX:2018akb, Aprile:2018dbl, XENON:2020kmp, LZ:2019sgr, PandaX-4T:2021bab} can. These detectors maintain extremely low background levels as their primary goal is to detect dark matter caused nuclear recoils. Because of it, it is not necessary to register both interaction products to register a neutrino interaction. While the completed detectors~\citep{Aprile:2018dbl} can only match the upper limits from neutrino electron elastic scattering, the currently running and planned detectors~\citep{Aalbers:2016jon, XENON:2020kmp, LZ:2019sgr, PandaX-4T:2021bab} can improve these limits approximately by two orders of magnitude~\citep{Strigari:2009bq, Suliga:2021hek}; see Figure~\ref{fig:DSNB-limits-nux}. In addition, depending on the exact exposure and the status of the uncertainty of the low energy atmospheric neutrino flux, this could potentially lead to detection.

Another potential channel for detecting the $\nu_x$ component of the DSNB is the elastic scattering of neutrinos on protons in liquid scintillator detectors such as JUNO~\citep{Tabrizi:2020vmo}.
This detection channel, however, has considerable challenges, a vast number of background events originating from solar neutrinos, radioactive decays of detector material, and cosmogenic backgrounds~\citep{JUNO:2015zny}.

\section{Conclusions}
The diffuse supernova neutrino background is a guaranteed neutrino flux that encodes information about the entire core-collapse supernova population in the observable universe. 
Its detection can serve as an independent indirect measurement of the core-collapse supernova rate, the fraction of black-hole-forming supernova progenitors, and the average neutrino flux per progenitor.
 Due to the unknowns connected to the neutrino flavor evolution in the dense environments and potential physics beyond the Standard Model in the neutrino sector, as well as uncertainties of the astrophysical parameters, to extract the most information, it is vital to detect diffuse supernova neutrino background in all flavors.
The prospects for detecting the $\bar\nu_e$ and $\nu_e$ are optimistic, especially now that the gadolinium-enriched Super-Kamiokande detector has already started seeing over 2$\sigma$ hints~\cite{SK-slides-2024}. 
New strategies for detecting the non-electron flavor component are needed.

\begin{ack}[Acknowledgments]%
The author would like to thank Baha Balantekin for the invitation to participate in this review. This work was supported by the Department of Energy grant No. DE-AC02-07CHI11359: \emph{Neutrino Theory Network Program}. Preprint number: FERMILAB-PUB-25-0745-T.
\end{ack}

\bibliographystyle{Numbered-Style} 
\bibliography{References}

\end{document}